\documentclass[sn-mathphys-num]{sn-jnl}

\usepackage{graphicx}%
\usepackage{multirow}%
\usepackage{amsmath,amssymb,amsfonts}%
\usepackage{amsthm}%
\usepackage{mathrsfs}%
\usepackage[title]{appendix}%
\usepackage{textcomp}%
\usepackage{manyfoot}%
\usepackage{booktabs}%
\usepackage{algorithm}%
\usepackage{algorithmicx}%
\usepackage{algpseudocode}%
\usepackage{listings}%
\usepackage{array}
\usepackage{makecell}
\usepackage{colortbl}
\usepackage{longtable}
\usepackage{adjustbox}
\usepackage{pdflscape}
\usepackage{placeins}
\usepackage{color,soul}
\usepackage{hyperref}
\usepackage{svg}
\usepackage{moresize}
\usepackage{booktabs}

\begin{document}

\title[Article Title]{Large-scale artificial intelligence with 41 million nanophotonic neurons on a metasurface}


\author[1]{\fnm{Mingcheng} \sur{Luo}}

\author[2]{\fnm{Meirui} \sur{Jiang}}

\author[3]{\fnm{Bhavin} \sur{J. Shastri}}

\author[4]{\fnm{Nansen} \sur{Zhou}}

\author[1]{\fnm{Wenfei} \sur{Guo}}
\author[1]{\fnm{Jianmin} \sur{Xiong}}

\author[1]{\fnm{Dongliang} \sur{Wang}}

\author[4]{\fnm{Renjie} \sur{Zhou}}

\author[1]{\fnm{Chester} \sur{Shu}}

\author[2]{\fnm{Qi} \sur{Dou}}

\author*[1]{\fnm{Chaoran} \sur{Huang}}\email{crhuang@ee.cuhk.edu.hk}

\affil*[1]{\orgdiv{Department of Electronic Engineering}, \orgname{The Chinese University of Hong Kong}, \orgaddress{\street{Shatin}, \city{New Territories},  \state{Hong Kong}}}

\affil[2]{\orgdiv{Department of Computer Science and Engineering}, \orgname{The Chinese University of Hong Kong}, \orgaddress{\street{Shatin}, \city{New Territories},  \state{Hong Kong}}}

\affil[3]{\orgdiv{Centre for Nanophotonics, Department of Physics, Engineering
Physics \& Astronomy}, \orgname{Queen's University}, \orgaddress{\street{64 Bader Lane}, \city{Kingston K7L3N6 ON}, \country{Canada}}}

\affil[4]{\orgdiv{Department of Biomedical Engineering}, \orgname{The Chinese University of Hong Kong}, \orgaddress{\street{Shatin}, \city{New Territories},  \state{Hong Kong}}}


\abstract{Conventional integrated circuits (ICs) struggle to meet the escalating demands of artificial intelligence (AI). This has sparked a renewed interest in an unconventional computing paradigm: neuromorphic (brain-inspired) computing. However, current neuromorphic systems face significant challenges in delivering a large number of parameters (i.e., weights) required for large-scale AI models. As a result, most neuromorphic hardware is limited to basic benchmark demonstrations, hindering its application to real-world AI challenges. Here, we present a large-scale optical neural network (ONN) for machine learning acceleration, featuring over 41 million photonic neurons. This system not only surpasses digital electronics in speed and energy efficiency but more importantly, closes the performance gap with large-scale AI models.  Our ONN leverages an innovative optical metasurface device featuring numerous spatial modes. This device integrates over 41 million meta-atoms on a 10 mm$^2$ metasurface chip, enabling the processing of tens of millions of weights in a single operation. For the first time, we demonstrate that an ONN, utilizing a single-layer metasurface, can match the performance of deep and large-scale deep learning models, such as ResNet and Vision Transformer, across various benchmark tasks. Additionally, we show that our system can deliver high-performance solutions to real-world AI challenges through its unprecedented scale, such as accelerating the analysis of multi-gigapixel whole slide images (WSIs) for cancer detection by processing the million-pixel sub-image in a single shot.  Our system reduces computing time and energy consumption by over 1,000 times compared to state-of-the-art graphic processing units (GPUs). This work presents a large-scale, low-power, and high-performance neuromorphic computing system, paving the way for future disruptive AI technologies.}




\maketitle

\section{Introduction}\label{sec1}

Neural networks (NNs) have become a powerful tool across various scientific and technological domains, triggering transformative shifts in fields such as drug discovery, image processing, autonomous vehicles, and medical diagnostics [1]. However, the growing complexity of problems is causing the cost of training and inferring NN models to double every two months [2], significantly outpacing the advancements in complementary metal-oxide-semiconductor (CMOS) circuits. Concurrently, there is an urgent need to minimize latency in training and inference processes, driven by the rise of time-sensitive applications such as navigation of self-driving cars, robotics, and real-time analytics for healthcare and surgery. While multi-core and multi-processor architectures can address the limitations of single processors, the rising demand for data movement creates interconnect bottlenecks, adversely impacting both computing time and energy consumption [3-6].

To address these challenges, an unconventional computing paradigm, neuromorphic computing, has recently gained traction [7-13]. In neuromorphic computing systems, neural network weights (connections between nodes) are stored in non-volatile memory and co-located with the computational elements. This configuration alleviates the data movement bottleneck, substantially improving computing speed and energy efficiency. Optical neural networks (ONNs) are particularly promising for neuromorphic computing due to the high degree of parallelism inherent in light waves [7, 10, 11, 14, 15]. By harnessing this parallelism, ONNs can execute linear operations, such as matrix multiplications, in a single operation. As a result, for tasks where computational complexity scales as $\mathcal{O}({N^{2}})$, both the computational time and the energy cost can be reduced to $\mathcal{O}{(1})$. Consequently, ONNs offer significant energy efficiency and speed advantages, especially when handling sizable weight and input numbers.

Despite the theoretical potential of ONNs, current implementations face challenges in scaling up in terms of parameter numbers. Two-dimensional (2D) integrated ONNs offer high computing speed through high-speed optoelectronic devices [16-38] but are constrained by the large footprint of optical devices, the insertion loss, and the complexities involved in controlling these components. These limitations restrict the maximum number of components demonstrated to around a few thousand [31, 39, 40]. By transitioning to 3D free-space optics, spatial parallelism in optics allows for a significant increase in the number of parameters, surpassing electronic systems [11, 14, 41-44]. Experimental demonstrations have successfully realized approximately 100,000 scalar multiplications [41, 45-50]. However, the parameter count still falls short of theoretical expectations and is significantly smaller than current AI models, which may have millions to billions of parameters [51, 52]. The challenge in further scaling 3D ONNs lies in the burden of training and accurately implementing a large number of physical parameters. A slight misalignment at the sub-wavelength scales can lead to significant reductions in accuracy. These errors further accumulate as the width (the number of nodes at each layer) of the ONN increases. Furthermore, the lack of efficient optical nonlinearity prevents ONNs from extending their depth. As a result, the system performance of ONN, such as classification accuracy, exhibits a considerable performance gap compared to state-of-the-art AI models.

Here, we present a 3D ONN based on optical metasurfaces, achieving competitive performance with large-scale AI models. Optical metasurfaces are highly compact free-space optic devices composed of numerous sub-wavelength meta-atoms arranged in a precise pattern on a 2D plane. Each meta-atom can independently and locally modulate the phase and amplitude of light [53-57]. The modulated wavefront serves as a source of secondary spherical waves, shaping the new wavefront in the next moment. Consequently, meta-atoms function as optical neurons that are fully-connected to adjacent layers [49-50]. Advanced fabrication technology can produce meta-atoms with a density exceeding 10 billion per cm$^2$,  comparable to the transistor density in advanced CMOS processors. Therefore, metasurfaces can offer over a billion optical neurons within a single 1 cm$^2$ chip, providing unparalleled computing parallelism in a passive way. However, despite its theoretical potential, this approach faces challenges similar to those encountered in other 3D ONNs, due to the difficulties in precisely fabricating and implementing large-scale arrays of meta-atoms. As a result, both the scalability and overall system performance remain significantly limited. In addition, current methods cannot actively tune metasurface at scale, making metasurface-based ONNs mostly single-tasked. 

In contrast to prior work, we experimentally demonstrate a large-scale and versatile ONN that integrates over 41 million meta-atoms in a single metasurface chip. This setup represents the largest neuron capacity ever shown in an experimental setting. Unlike other ONNs, which are trained to a specific task, we adopt a new computing framework based on random projection. Random projection operates as a universal kernel machine [58, 59], offering broad generalizability across a wide range of tasks. Assisted by a highly compact and programmable electronic NN at the backend, we can overcome the non-reconfigurability in metasurfaces, making the overall system trainable, versatile and achieve best-in-class performance across different tasks. Furthermore, random projection only requires the transmission matrix to follow a Gaussian distribution without the need for precise design of each individual meta-atom. This flexibility enables our system to scale without being constrained by fabrication and implementation errors, allowing the meta-ONN to expand to arbitrary widths, depths, and highly complex neural network models.

Our theory and experiment shows that when the neuron count reaches the scale of tens of millions, our metasurface-based ONN (meta-ONN) behaves like an infinitely wide NN. A single layer of ONN, together with a digital NN with fewer than 10,000 parameters, can achieve performance levels that are competitive with large NN models like Residual Neural Network (ResNet) [89] and Vision Transformer [52] with millions to billions of parameters. Furthermore, we also find that when initializing the meta-atoms with a Gaussian distribution, even without any training process, our meta-ONN can achieve comparable and satisfactory performance to a trained network.

To demonstrate the versatility and high performance of our approach, we showcase a range of applications, including image classification and detection. All tasks, using a single metasurface layer and a compact digital model with fewer than 10,000 parameters, achieve performance far surpassing current ONNs and rivaling deep, large-scale AI models. To illustrate the exceptional scalability of our scheme, we process high-resolution medical images with over a million pixels. Additionally, we implement a recurrent neural network (RNN) with optical metasurfaces for human action recognition, achieving an impressive accuracy of 99.1\%. This highlights the system’s ability to scale to deep layers without being constrained by physical fabrication errors. Leveraging these remarkable advantages, we further demonstrate that our system addresses real-world challenges beyond the reach of existing ONNs to accelerate the analysis of multi-gigapixel whole slide images (WSIs) for cancer detection by processing million-pixel sub-images in a single shot.

By conducting over 99.995\% of computations in the optical domain with a passive metasurface chip, we achieve an energy efficiency of 241 TOPS/W (\textcolor{blue}{Supplementary Note 21}). This figure includes the power consumption of peripheral circuits for optical signal generation and detection. Our meta-ONN not only surpasses digital electronics in speed and energy efficiency, but it also matches the performance of state-of-the-art large AI models in accuracy and versatility. Our work offers a novel and high scalable pathway for enabling large-scale AI computing with optical systems.

\section{Results}\label{sec2}
\subsection{System description and working principle}\label{subsec2_1}
The implementation of the proposed meta-ONN is illustrated in Fig.1a. The meta-ONN consists of a dielectric metasurface containing massive silicon cylindrical meta-atoms arranged in the two-dimensional plane with a fixed period. Each meta-atom controls the transmissive and reflective phase and amplitude of incident light at a subwavelength scale. These modulation values are determined by the interference between electric and magnetic dipole resonances within each meta-atom, which can be adjusted by varying the meta-atom diameter [60-63]. The incident beam, carrying encoded input information such as images, is reflected by the metasurface. According to the Huygens–Fresnel principle, the wavefront, after being modulated by each meta-atom, acts as secondary spherical waves and spreads out in all directions. The resulting wavefront at the following plane is formed by the combination of all these secondary wavelets, with each wavelet contributing to the overall shape of the wavefront [49, 50]. Consequently, each meta-atom can be regarded as an optical neuron, fully connected to the image sensor array. Since the number of pixels in the detector (480,000 pixels in our experiments) is significantly smaller than the number of meta-atoms (41 million), the optical field at the receiver is effectively downsampled through sum pooling, where neighboring elements are summed together. This process forms an extremely wide single-layer neural network with 480,000 $\times$ 41 million weights, though the actual degrees of freedom for tuning are limited to the 41 million meta-atoms. An optical lens is placed between the metasurface and the image sensor array to perform a Fourier transform, which, as we will show later, enhances the network’s ability to extract high-frequency features. The image sensor array then applies a square function to the pooled signal, providing the nonlinearity of the NN. The output of each sensor pixel is $ \tilde U_{k}=[\sum_{i=k\cdot R}^{(k+1)\cdot R - 1} \mathcal{F}({\textbf{\emph{H}}}_{d}{\textbf{U}}_{0}\otimes\textbf{\emph{W}})[i,:]]^{2}$, where $\textbf{U}_{0}$ is the input image, $\textbf{\emph{W}}$ is the transmission matrix of the metasurface, $\textbf{\emph{H}}_{d}$ is the function for optical diffraction, $\mathcal{F}(\cdot)$ is the operator of Fourier transformation, $N$ is the number of meta-atoms, $M$  is the number of camera pixels, and $R$ is the ratio of these two quantities ($R=N/M$) which is also the downsampling ratio. This process can be treated as first projecting the original input into an $N^{2}$ dimension space through the above formula and performing sum pooling to downsample the high-dimension space to a $M$ dimension space (Supplementary Note 3).  This makes our system also function as a universal kernel machine, mapping input images into a $N^{2}$ high-dimensional Fourier feature space [59].

When designing the transmission of each meta-atom, we adopt a novel strategy that ensures scalability to 41 million meta-atoms without being constrained by physical fabrication errors. Our approach is inspired by the Neural Tangent Kernel (NTK), a theoretical framework for analyzing the behavior of infinitely wide neural networks [64-70]. NTK theory suggests that when weights are initialized with a Gaussian distribution, they have minimal change during training, meaning that Gaussian-initialized weights, even without training, are already close to the global minimum [64, 67]. Therefore, instead of training the metasurface for specific tasks, we design the meta-atoms to directly follow a Gaussian distribution. This approach requires only the transmission matrix to match a Gaussian distribution, eliminating the need for precise tuning of each individual meta-atom. As a result, our system gains remarkable flexibility, enabling the meta-ONN to scale without being limited by fabrication and implementation errors, and allowing expansion to arbitrary widths, depths, and highly complex neural network models.

Our fabricated metasurface consists of 6,400 $\times$ 6,400 = 41 million silicon meta-atoms, of which diameter is designed to be varied from 100 to 400 nm at a unit cell$'$s period of 500 nm, as shown in Fig.2c. These meta-atoms are designed to provide a complex-valued transmission matrix randomly sampled from a Gaussian distribution. The standard deviation of this Gaussian distribution is designed as large as 0.4 $\pi$ to ensure a higher degree of randomness and thus support high performance (see metasurface design in Methods). The measured complex-valued transmission matrix of the metasurface is shown in Fig.2d. The insertion loss of the whole optical system is measured as 7.2 dB, which can be reduced by using low-loss materials, for example, sapphire and titanium dioxide [71, 73].

\begin{figure}[H]
\centerline{\includegraphics[trim={0.15cm 0cm 0.1cm 0.1cm},clip,scale=0.68]{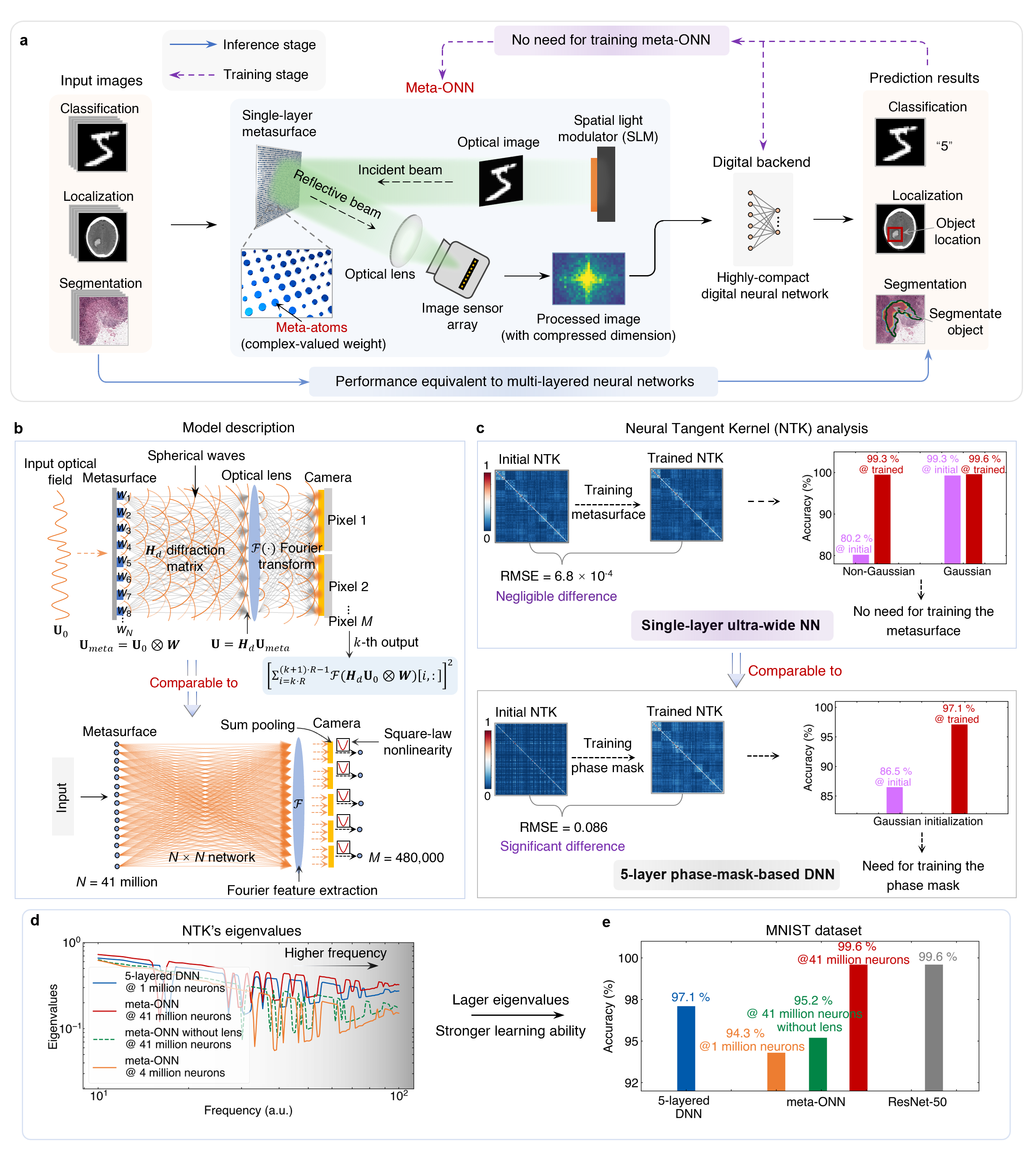}}
\caption{The metasurface-based optical neural network (meta-ONN). (a) Schematic of the meta-ONN. The optical image generated by the SLM is projected onto a single-layer metasurface consisting of massive cylindrical silicon meta-atoms. An optical lens then collects the reflected optical image by the metasurface. Following that, the optical field at the focusing plane is captured by an image sensor array. Lastly, the captured image is fed into a digital neural network to produce the prediction result. (b) The model description of meta-ONN. (c) The NTK analysis of the meta-ONN (upper) and 5-layered phase-mask-based DNN (bottom) during the training stage. (d) The eigenvalues of the trained NTKs over the frequency range for various ONNs. (e) The comparison of the accuracies obtained from various ONNs without training the optical layer. }
\label{fig1}
\end{figure}

Our method shares similarities with optical computing systems that leverage random scattering [10, 74-78] and extreme learning machines (ELMs) [79, 80], and are closely related to reservoir computing [81-85]. However, our system has three critical differences leading to its excellent performances compared to other systems. First, the introduction of a metasurface provides, in practice, an infinitely wide layer to encode the input into an extremely high-dimensional space over (41 million)$^{2}$, a scale never demonstrated in prior systems.  Second, using metasurfaces provides full controllability of each entry (meta-atom) to optimize the projection matrices by ensuring that the NN is initialized with Gaussian distribution, as opposed to solely relying on the random nature of a physical system including random scattering media and multi-mode optical fibers [74, 79], where the optical behavior cannot be precisely engineered to achieve the desired Gaussian distribution, ultimately degrading performance. Third, we use an optical lens to provide a Fourier transform, which is critical for the system to learn high-frequency functions [70]. These distinctions enable a single-layer metasurface to rival cutting-edge NN models with many nonlinear layers, such as ResNet and Vision Transformer—a performance that has never been realized in other ONN systems.

With these unique distinctions, our system exhibits properties not observed in current ONNs, and we validate these advantages using NTK theory by analyzing the eigenvalues of NTK across different frequencies [70, 86], where higher eigenvalues indicate greater learning ability. We further demonstrate these advantages through MNIST digit classification as a benchmark. The details of NTK simulations are provided in \textcolor{blue}{Supplementary Note 4}.

First, with 41 million meta-atoms, a single-layer NN exhibits behavior equivalent to, or even surpassing, that of a multi-layered neural network. To illustrate this, we calculate the NTK for three systems: (1) a single-layer optical ONN with 41 million nodes (our proposed system), (2) a single-layer ONN with 4 million nodes, and (3) a multi-layer diffractive neural network (DNN) with the setting demonstrated in the reference [41], as shown in Fig.2c. The results show that when the neuron count reaches 41 million, the single-layer ONN matches and even exceeds the performance of a multi-layered ONN. Notably, a single-layer ONN achieves an MNIST classification accuracy of 99.6 \%, outperforming the multi-layered ONN. Moreover, multi-layered ONNs often face significant challenges related to the precise alignment of different layers, which can degrade overall system performance during implementation.

Second, when the neuron count reaches 41 million and the initial weights are distributed according to a Gaussian distribution, the NN can achieve comparable and satisfactory performance to a trained network, even without any training process. Fig.2c shows the changes in the NTK during training. When the neuron count reaches 41 million and the initial weights follow a Gaussian distribution (as in our system), the NTK remains nearly unchanged before and after training. This indicates that our system, even without training, the NN can provide a matched performance with that with training. This behavior aligns with NTK theory for infinitely wide NNs [64, 67]. We have experimentally achieved a high accuracy of 99.3 \% under the MNIST task using one training-free metasurface chip and only 3,000 trained digital weights, surpassing the current ONNs, as shown in Extended Table 1. The experimental details for the MNIST task are provided in \textcolor{blue}{Supplementary Note 12}. This property is valid only when the weights are initialized with a Gaussian distribution and with 41 million nodes, as shown in Fig.2c. In contrast, when the neuron count is limited, as in the case of a 5-layer diffractive NN with a total of 1 million neurons, training becomes essential for improving accuracy. This training-free neural network is generic rather than task-specific, allowing a compact digital layer to be attached to ensure both versatility and high performance.

Third, our system includes a lens that performs Fourier feature mapping [87, 88] within the optical domain. This Fourier feature mapping layer enhances the system’s ability to learn high-frequency features [70], and therefore provides even better performance compared to multi-layered NN. As shown in Fig.2d, incorporating the lens significantly increases the eigenvalues, particularly in the high-frequency region. This suggests that the lens enhances the ability to learn high-frequency components of the information. Consequently, accuracy is improved accordingly, as shown in Fig.2e.

\subsection{Rival deep NNs with a single-layer metasurface}\label{subsec2_2}
We begin by conducting machine vision tasks to demonstrate the high performance, versality, and high scalability of our system. For benchmarking, we compared the performance of our meta-ONN with three benchmarking large-scale deep learning models: ResNet-50 [89], a classical 50-layered CNN with approximately 23.5 million parameters; the Segment Anything model (SAM) [90], a cutting-edge large promotable segmentation model with 93.7 million parameters; and Vision Transformer (ViT) [52], a transformer encoder model for image classification with more than 85.8 million parameters (\textcolor{blue}{Supplementary Note 18}). For each task, the input images are generated using a SLM, as shown in Fig.2a. These images are processed by the metasurface and then collected by an optical lens before being detected by a CMOS digital camera. The detected digital image is downsampled, with the downsampling ratios being task-dependent. The same metasurface chip is used for all the tasks. Only the digital neural network at the backend is trained for different applications. 

\textit{COVID-19 Radiography}: we further explore the applications of our meta-ONN for practical use by applying it to the COVID-19 Radiography dataset [91] as shown in Fig.2e. This dataset includes over 20,000 chest X-ray (CXR) images covering normal and COVID-19 positive cases, each with 299 $\times$ 299 pixels. Using the same metasurface chip and experimental setup as with CIFAR-10, optical images of size 299 $\times$ 299 are generated by the SLM, processed by the meta-ONN, and subsequently detected. The detected image is dramatically downsampled to only 12 $\times$ 16 pixels, followed by a highly compact regression network with only 192 weights at the digital backend to produce the binary classification results for normal and COVID-19 cases. The accuracy, precision, sensitivity, and specificity obtained from the classification results are 98.0\%, 96.8\%, 99.2\%, and 96.9\% using only 192 digitally trained weights, respectively. The experimentally obtained accuracy of meta-ONN up to 98.0\%, outperforms previously demonstrated simulation and experimental ONNs [76, 79], as shown in Fig.2k. This accuracy is competitive against ResNet-50 (97.0\%) and SAM (97.2\%) and slightly lower than ViT (99.0\%). Nevertheless, our meta-ONN can compress the digital model by a factor of 4.9 $\times$ 10$^5$ compared to ResNet-50 and 1.8 $\times$ 10$^6$ compared to ViT, signifying a remarkable reduction of computing time by 1.2 $\times$ 10$^5$  times at the training stage and energy consumption by 1,900 times at the inference stage. The computing time includes the time required for SLM response, the free-space propagation of the light, camera response, and executing a digital computer, respectively. The energy consumption includes the energy consumed by the laser source, SLM, camera, and digital computer, respectively. More details about the time and energy consumption benchmarks can be found in \textcolor{blue}{Supplementary Notes 19 and 20}.

Our investigation further shows that the substantial number of optical weights unique to the metasurface device is the key factor contributing to the high accuracy. The accuracy increases from 83.0\% to 98.0\% as the number of optical weights increases from 4.2 million to 41 million as shown in Fig.2h (see details about experiments in \textcolor{blue}{Supplementary Note 7}). To gain deeper insights into the data, we employ t-distributed stochastic neighbor embedding (t-SNE) to visualize the data after being processed by the meta-ONN, where the original images are transformed into a lower-dimensional space by the metasurface. As shown in the inset graphs of Fig.2h, the separation between different classes becomes highly distinct as the weight number increases substantially to 41 million.

Encouraged by the performance we have achieved, we delve into more challenging applications that have never been demonstrated by ONNs. These tasks showcase our meta-ONN's ability to process high-resolution images and tackle more challenging medical diagnostic tasks.

\textit{Thoracic diseases detection}: The NIH ChestX-Ray8 dataset [92] comprises high-resolution (1,024 $\times$ 1,024 pixels) front-view X-ray images covering eight different thoracic diseases as shown in Fig.2f. We demonstrate that these commonly occurring thoracic diseases can be detected by our meta-ONN with an average accuracy of 85.4\% as shown in Fig.2i, followed by minimal digital processing using 9,600 digitally trained weights. The achieved accuracy is comparable to results from ViT (85.8\% accuracy) employing 85.8 million digitally trained weights. Additionally, this task highlights our meta-ONN's ability to process high-resolution images without any pre-processing, utilizing a pixel resolution of 1,024 $\times$ 1,024, the highest demonstrated in ONNs to date to our best knowledge.

\textit{Intracranial Hemorrhage Detection}: Intracranial hemorrhage (ICH) is a critical medical condition that requires rapid and intensive treatment to prevent further damage and to improve patient outcomes [93]. However, the timely detection of ICHs is often delayed due to a lack of prompt access to radiologists who read the scans. Here, we leverage our meta-ONN to expedite ICH detection in CT images. Our experiments solve two tasks. In the first task, we detect the existence of hemorrhages in CT images, achieving an accuracy of 98.8\%, a precision of 98.5\%, a sensitivity of 99.2\%, and a specificity of 98.3\%. In the second task, we identify the bleeding point location with the meta-ONN. We first downsample the optical images after meta-ONN to 15 $\times$ 20 pixels. These downsampled images serve as the input to a 6-layered fully connected neural network at the digital backend. The output is a vector with 4 elements, each element representing the horizontal and vertical position coordinates, the width, and the height of a rectangular box. The box indicates the bleeding region in the CT images, as shown in Fig.2j. Intersection over Union (IoU) is obtained from the predicted box and the true box to evaluate the performance of locating the bleeding region. Here, IoU is the ratio of the overlapped area over the united area of the two boxes. Our meta-ONN achieves an IoU of 0.61 for 3 different bleeding positions, meaning that the predicted bounding box indicating the bleeding location is well aligned with the ground truth box. In comparison, the averaged IoU obtained by ResNet-50 is 0.64.

An important discovery implied by the results of our meta-ONN across diverse vision tasks is its capacity of using only a single-layer metasurface to compete against cutting-edge NN models with tens of millions of parameters. We experimentally validate in different tasks that our superior performance is credited to the massive parameters uniquely offered by the metasurface, as demonstrated in Fig.2k to Fig.2m. While a small NN is still necessary at the digital backend, its size can be highly compressed to up to a factor of 10$^5$. Furthermore, it is worth noting that the remaining small NN at the backend can be readily implemented in analog devices, enabling the realization of an all-analog system.

\begin{figure}[H]
\centerline{\includegraphics[trim={0.1cm 0.1cm 0.2cm 0.2cm},clip,scale=0.67]{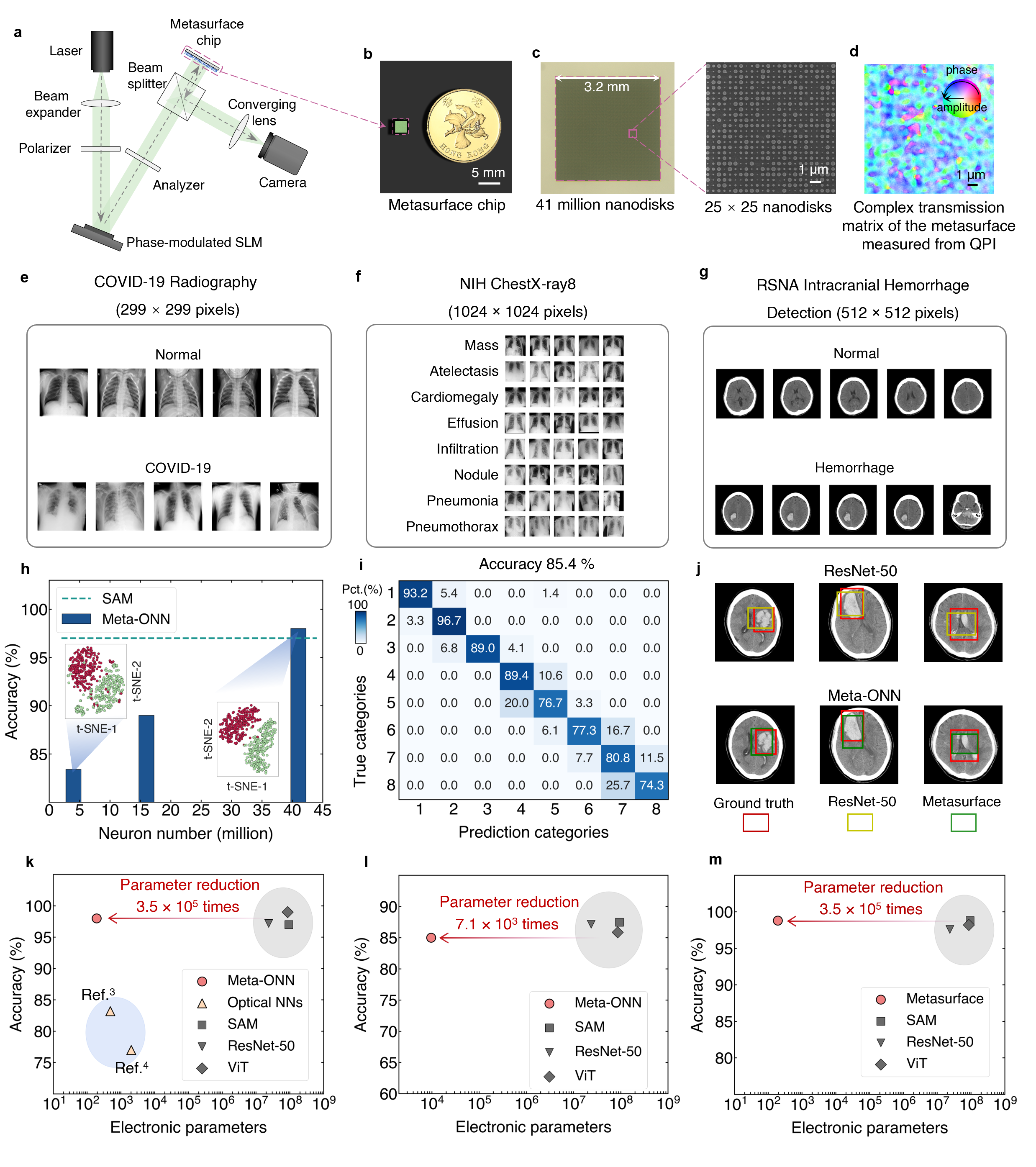}}
\caption{Experimental results of the meta-ONN for three benchmark tasks. (a) Schematic of the experimental setup of meta-ONN for the benchmark task. (b) Image of the metasurface chip compared with a Hong Kong dollar coin. (c) Optical microscope image of the fabricated metasurface chip with a compact area of 3.2 $\times$ 3.2 mm$^{2}$ (left graph) and the scanning electron microscope (SEM) image of a zoom-in metasurface region containing 25 $\times$ 25 meta-atoms (right graph). (d) The measurement result of quantitative phase imaging (QPI) of a zoom-in region of the metasurface chip. The details about the QPI measurement are provided in \textcolor{blue}{Supplementary Note 5}. In the color wheel, the color represents the phase modulation coefficient, and the brightness represents the amplitude modulation coefficient. (e)–(g) Illustration of the dataset images of the COVID-19 Radiography (e), the NIH ChestX-ray8 (f), and RSNA Intracranial Hemorrhage Detection tasks (g). (h) The accuracy versus optical neuron number of the meta-ONN for the COVID-19 Radiography. The dashed line represents the accuracy of SAM as a comparison. The inset graphs with colorful scatters represent the results of the t-SNE analysis. (i) The confusion matrix of the prediction result of the NIH ChestX-ray8. (j) Bleeding regions inside the brain predicted by the meta-ONN (bottom graph) and digital ResNet-50 (upper graph). (k)–(m) Comparison of accuracy and electronic parameters of the meta-ONN with other optical approaches [76, 79, 94, 95] and digital models for the COVID-19 Radiography (k), the NIH ChestX-ray8 (l), and RSNA Intracranial Hemorrhage Detection tasks (m).}
\label{fig2}
\end{figure}

\subsection{Scalable to complexed NN models for high-accuracy video sequence processing}\label{subsec2_3}

Leveraging random projection, we can confidently increase the depth of the optical NN without concerns about physical errors accumulating layer by layer. This allows our system to tackle more complicated AI tasks. Here we construct a recurrent neural network (RNN) using metasurface-based ONN and apply it in video processing. Serving as a hidden layer, the ONN is recurrently connected to form an RNN [46, 85]. This meta-RNN takes a sequence of image frames from a video as input, as shown in Fig.3a. Each frame is processed by the meta-RNN, with the output being downsampled and read out from the camera. Subsequently, this output is combined with the hidden state from previous frames to derive the hidden state of the current frame. Finally, the output is connected to a logistic regression-based classifier. The meta-ONN is employed for human action recognition tasks on the KTH dataset [96], which comprises six types of human actions across four different scenarios.

The evaluation of the performance of the model involves two types of action classification. One type identifies actions jointly determined by sequences of frames, defined as action accuracy. The action is determined from the percentage of votes for all actions in each testing video sequence. The action with the maximum votes is the predicted action in a video sequence. The other type identifies the action indicated by each individual frame, defined as frame accuracy. With 9,180 trained weights, our meta-RNN obtains the frame accuracy of 97.5\% and action accuracy of 99.1\%, as shown in Fig.3b and Fig.3c, respectively. The training time to realize such a high accuracy is only 4.01 seconds using NVIDIA RTX 3090. The obtained accuracy exceeds those from digital NNs using long-short term memory (action accuracy of 90.7\%) and the state-of-the-art ONNs (action accuracy of 96.3\%). Additionally, unlike other ONNs that typically need preprocessing of the frames of the action sequences by pre-trained CNN for human segmentation, our meta-ONN can directly process videos without the need for such preprocessing steps. Consequently, our meta-ONN is capable of processing videos at high speed, achieving 1,968 frames per second.

We further utilize this task to demonstrate the capability of our ONN to rapidly recover from external perturbations and its potential for real-time adaptable learning. To show this property, we intentionally move the metasurface by 10 $\mu$m creating the axial misalignment after the digital NN is trained. The misalignment is the 20 times of the light wavelength. This axial misalignment results in a change in the modulation matrix of the metasurface, which may lead to the degradation of the accuracy by 16.7\%. However, the meta-ONN is capable of quickly recovering from the error after re-training the single-layer digital NN as shown in Fig.3e. The accuracy is restored to 96.3\% after 45 epochs using only 1,920 training images, which takes only 234 ms. This study highlights the adaptability of the meta-ONN, even in dynamic or unpredictable environments and its potential for real-time adaptable learning. The experimental details can be found in \textcolor{blue}{Supplementary Note 8}. Computing very large matrices with optical fan-in allows for extremely low optical energy consumption. If the matrix size is sufficiently large, each multiplication requires a photon number far less than 1. Here, we demonstrate in Fig.3f that an average photon number of 0.078 per multiplication is sufficient to maintain human action classification accuracy over 95\%. This feature makes our system suitable for low-illumination environments. The experimental details can be found in \textcolor{blue}{Supplementary Note 9}.

\begin{figure}[H]
\centerline{\includegraphics[trim={0.2cm 0.1cm 0.2cm 0.25cm},clip,scale=0.570]{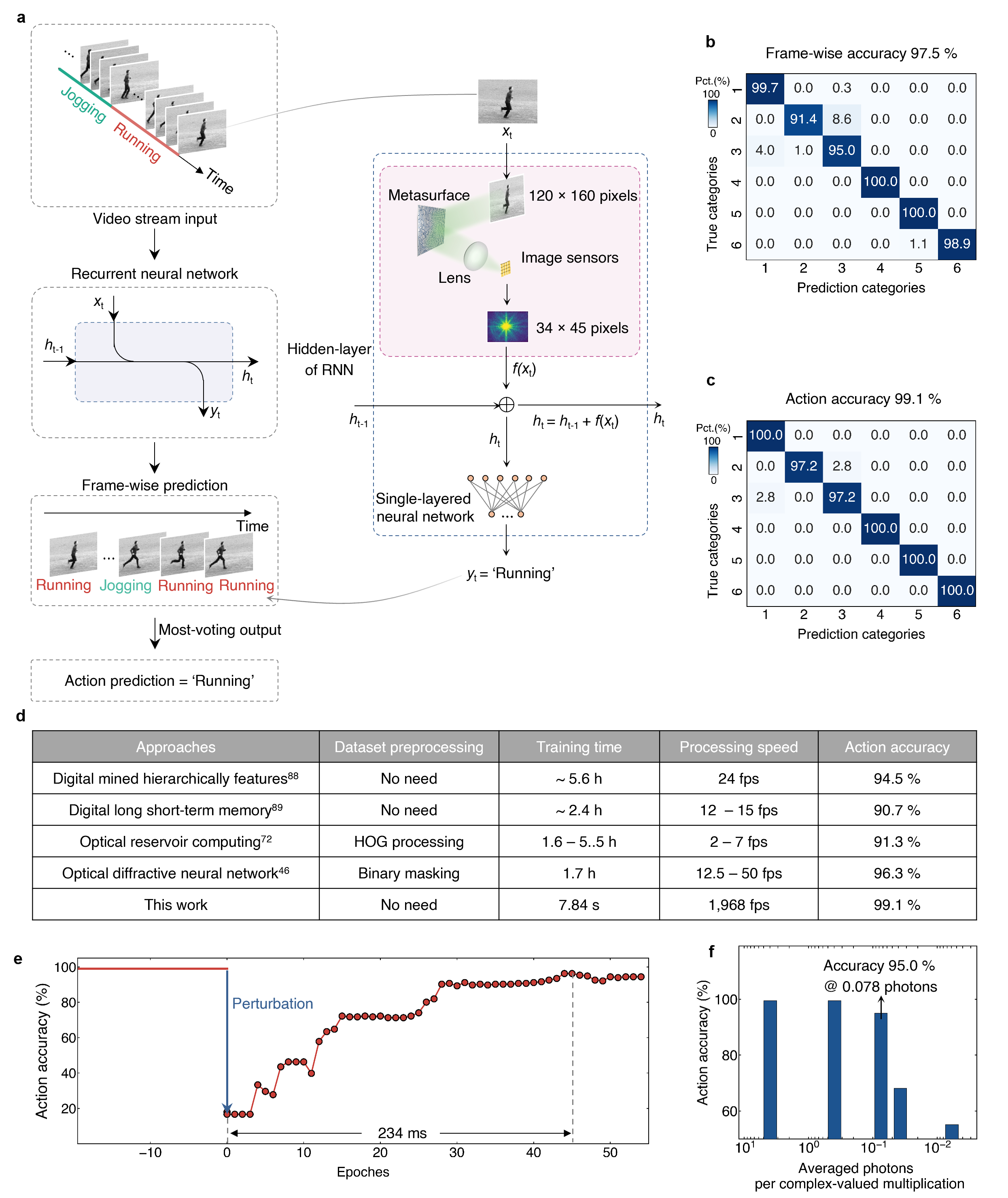}}
\caption{Experimental results of the meta-RNN for video-based human action recognition. (a) The working flow of the meta-RNN for video-based human action recognition. The action video is processed by the meta-RNN in a frame-by-frame way. Each action video contains 20 frames with a time interval of 80 ms. Firstly, the input frame $x_{t}$ with 120 $\times$ 160 pixels is encoded into the optical domain without any preprocessing. It is then processed by the meta-ONN to generate a compressed digital image $f(x_{t})$ with 34 $\times$ 45 pixels. The processed frame is fused with the output at the previous frame to form the output at the current frame, i.e., $h_{t} = h_{t-1} + f(x_{t})$. Finally, the output at the current frame is fed into a digital neural network to produce a frame-wise prediction result $y_{t}$. The action prediction result is the label that appears most frequently in all frame-wise results of the action video. (b) The confusion matrix of frame-wise prediction results. (c) The confusion matrix of action prediction results. (d) Performance comparison of the meta-ONN with the state-of-the-art digital and optical approaches [46, 77, 96, 98]. The training time and processing speed of the meta-RNN are the predicted ones based on current experimental results (see details in \textcolor{blue}{Supplementary Note 19}). (e) The ability of the meta-RNN in quickly recovering the accuracy after experiencing a hard perturbation. (f) The action accuracy versus averaged photons per complex-valued multiplication.}
\label{fig3}
\end{figure}

\subsection{Solving real-world challenge: accelerating the analysis of multi-gigapixel whole slide images for cancer diagnosis}\label{subsec2_4}

Leveraging on these advantages, we finally demonstrate the application of our meta-ONN in a real-world challenge, showcasing its distinct benefits in computationally intensive applications. For many diseases, particularly cancers, pathological diagnosis is the gold standard in clinical practice. The introduction of Whole Slide Imaging (WSI) scanners, which generate digitized pathology microscopic images, has revolutionized pathology image analysis. This technology enables computer-aided diagnostics, leveraging advanced deep-learning techniques to reduce the workload of pathologists and optimize the regional distribution of medical resources [99]. However, WSIs present a challenge for deep learning due to their extremely large size. With single slides containing multi-gigapixel images, efficient processing of these images is crucial for automated diagnosis.

In our work, we apply a meta-ONN to detect and localize breast cancer that has metastasized to nearby lymph nodes, a task of significant clinical importance but requiring substantial reading time from pathologists. Pathologists typically need to review thousands of megapixel photos from a single WSI exceeding 10 gigapixels. We use the CAMELYON16 dataset [100] for our study. To address the processing of extensive WSIs, each with dimensions of more than 2 billion pixels, we employ a patch-based framework. Initially, a preprocessing algorithm, Otsu algorithm [101], is used to separate the raw whole slide image into the useful foreground and the non-tissue background, resulting in a reduced total pixel of the whole slide image to be processed, as shown in Fig.4a. We then divide the large WSIs into smaller patches, each containing 1,000 $\times$ 1,000 pixels. These patches consist of 1,775 normal patches and 887 tumor patches. During the training phase, we convert the patches into optical images using the SLM. The modulated patch samples are processed by our meta-ONN and detected by the sensor array. Subsequently, the output from the sensor array is downsampled. The final step is to train a single-layered neural network with the patch samples to create a classifier capable of distinguishing between normal and tumor classes. The mean AUC is 96.0\% when the trained weight number is 140 and increases to 97.0\% at the weight number of 1,200, as shown in Fig.4b. The training process takes only 1.46 s, achieving a training accuracy of 95.1\%, as shown in Fig.4d.

Another WSI consisting of 2,030 unlabeled patches is used to test the performance of tumor tissue segmentation. Initially, these unlabeled patches are first processed using our meta-ONN. Subsequently, the processed patches are inferred with the trained single-layered NN to produce the tumor-positive probability. Finally, the probability heat map is generated by mapping the predicted probability of the patches to the raw WSI. As shown in Fig.4c, the resulting heat map demonstrates that our meta-ONN achieves accurate segmentation of three different tumor tissue regions from the billion-pixel-scale WSI, achieving an IOU of 0.60, which is comparable to that of SAM (0.63). More importantly, our meta-ONN exhibits an impressively fast inference time of 1.02 s per whole slide image (WSI), representing a significant reduction compared to SAM which requires 1.48 hours to analyze one WSI. This remarkable reduction in inference time allows our meta-ONN to diagnose more than 42,352 patients within a single 12-hour working day, while only 8 patients can be diagnosed using SAM in the same timeframe.

\begin{figure}[H]
\centerline{\includegraphics[trim={0.2cm 0.2cm 0.2cm 0.2cm},clip,scale=0.7]{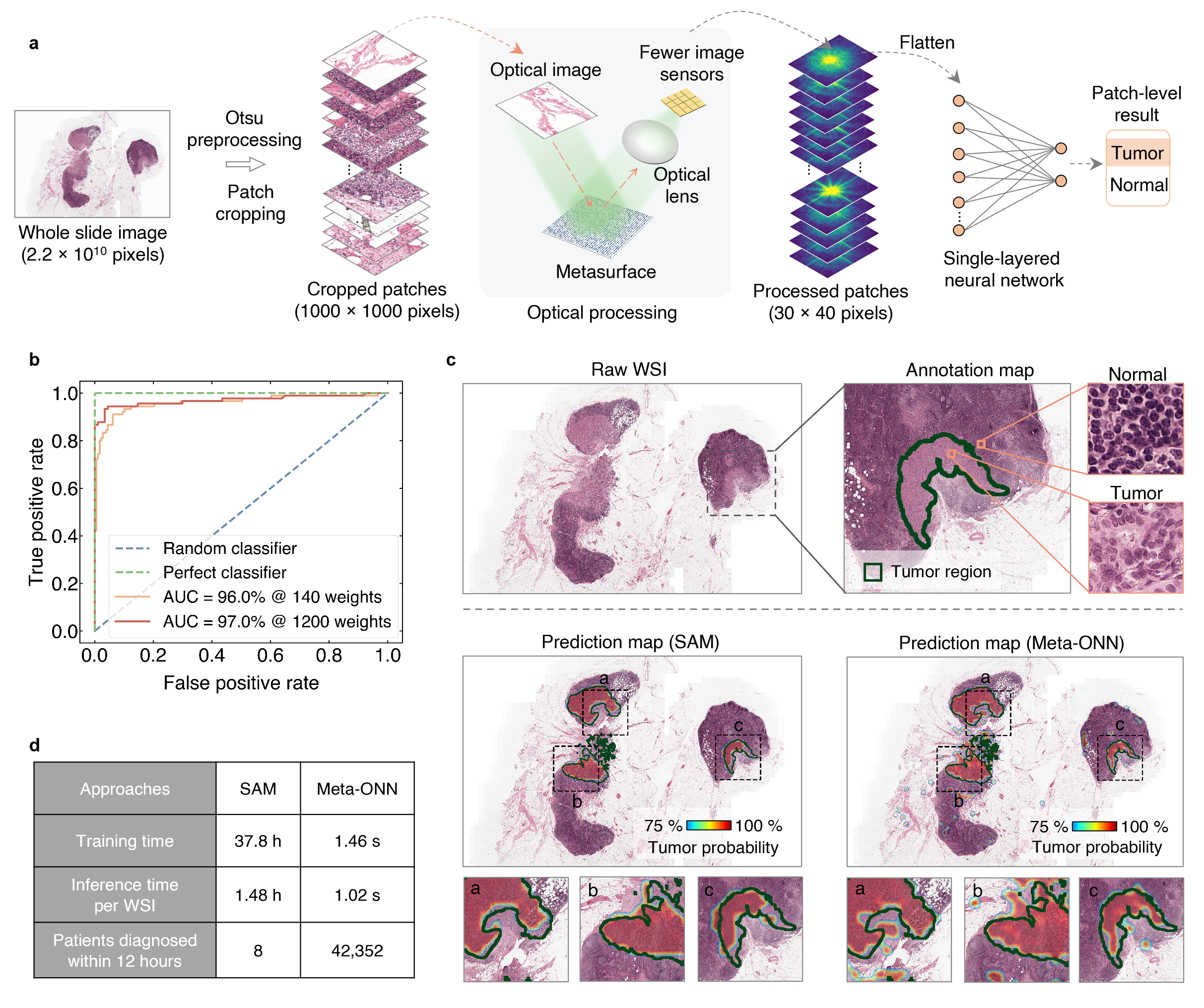}}
\caption{Experimental results of the meta-ONN for cancer diagnosis based on the whole slide image. (a) The working flow of the cancer diagnosis based on the whole slide image. The input WSI with over 2.2 $\times$ 10$^{10}$ pixels is preprocessed with the Ostu method and cropped into a series of patch images with 1,000 $\times$ 1,000 pixels. The patch images are processed by the meta-ONN in a single shot and compressed into 30 $\times$ 40 pixels. Finally, these processed images are fed to a highly compact digital neural network to generate the final prediction result of whether the tumor cell is detected in the image. (b) ROC curve of the patch-level prediction results during the training phase. (c) The heat map of the prediction probabilities of the meta-ONN and the state-of-the-art segmentation model (SAM). The inset graphs a, b, and c represent three different zoom-in regions of the WSI, respectively. (d) Comparison of the training time and inference time of the meta-ONN with SAM. The time consumption of the meta-ONN is the predicted one based on current experimental results (see the details in \textcolor{blue}{Supplementary Note 19}). }
\label{fig4}
\end{figure}

\section{Discussion and conclusion}\label{sec12}

We experimentally demonstrate a high-performance ONN that provides unprecedented accuracy, energy efficiency, and computing throughput. These exceptional performances are a result of realizing a true large-scale optical computing system by innovatively combining optical metasurfaces, a device fully leveraging the parallelism of optics, with random projection, a computing framework suitable for metasurface-based NNs to scale to arbitrary widths, depths, and complexity. An important observation from our experimental demonstration is that a single-layer optical metasurface chip, integrating 41 million optical neurons, can achieve competitive performance compared to state-of-the-art deep NNs such as ResNet and ViT. This result may alleviate current bottlenecks in extending ONNs to deeper layers, as realizing the nonlinearities between layers remains a significant challenge in ONN systems. The ability to realize large-scale AI models is often crucial for solving challenging tasks to meet downstream application requirements. Our work represents a significant advancement, moving ONNs beyond benchmark demonstrations to effectively address real-world challenges. The performance comparison of the meta-ONN with the state-of-the-art ONN systems is shown in \hyperref[Extended_table1]{Extended Table 1}.

By computing 41 million optical neurons in a single operation, our system experimentally realizes a computing throughout exceeding 4,700 Tera operations per second (see the calculation details in \textcolor{blue}{Supplementary Note 21}).  The neuron capacity and operation speed of the meta-ONN could be further improved by serval orders of magnitude. For example, the neurons can easily be scaled up to 1.6 billion using a larger footprint metasurface (20 $\times$ 20 mm$^2$) with a commercially available fabrication process. Vertical-cavity surface-emitting laser (VCSEL) arrays have demonstrated the ability to operate at the quantum-noise limit and achieve high speeds in the GHz range [102]. The computing throughput could be further enhanced by leveraging state-of-the-art high-speed VCSEL arrays.

Our system loads over 99.99\% computations on a passive optical metasurface. Although a small digital NN is still required at the backend, the time for training such as NN is greatly shortened from tens of hours to a few seconds. If high-speed SLMs and cameras are used, the training time could be reduced to only tens of milliseconds. This makes our system highly adaptable to dynamic or unpredictable environments that requires real-time learning abilities, such as robot-assisted surgery. Meanwhile, the inference time could easily be improved to the microsecond or even sub-microsecond level, which could enable a range of applications involving high-speed objects and requiring real-time decisions. Even shorter training and inference times could be achieved using delicate neural network accelerators realized with field-programmable gate arrays (FPGAs) and all-analog electronics [103, 104].

The sub-wavelength optical metasurface is much thinner and smaller than bulk optical devices. This allows for a highly compact integration with the CMOS sensors and thus a minimized system volume of our meta-ONN system. The system volume of a free-space-based ONN is defined as the area of the free-space optical device multiplied by the spacing distance between the optical device and the COMS camera. In the current experiment setup which uses a reflective metasurface, the reduction of volume is hindered by the use of a beam splitter, resulting in a system volume of 2,560 mm$^3$. In an improved design, the metasurface could be designed to work in a transmissive way by using a transparent substrate, for example, sapphire and quartz. The beam splitter is not needed. The distance between the metasurface and the converging lens could be shortened. In addition, the bulky converging lens could be replaced with a thin metalens with a high numerical aperture of 0.99 [105]. Considering that the metalens needs to fully cover the metasurface with a squared area of 3.2 $\times$ 3.2 mm$^2$, the diameter of the metalens is 4.5 mm and the focal length is 2.27 mm. The minimum distance between the metasurface and the lens can be equal to the metasurface size of 3.2 mm, which enables the all-to-all connection of optical weights during light diffraction. The spacing between the metasurface and the camera can be as small as 5.47 mm. As a result, the system volume could be extremely shrunk to 56 mm$^3$. which is more compact by 10$^2$ to 10$^4$ times than most state-of-the-art 3D free-space-based ONN systems [41, 44, 46, 48]. This compact volume allows the optical system to be fully integrated with the back-end digital processor.

One of the advantages of the meta-ONN is the ability to work under wide-spectral light conditions. The optical response of metasurfaces is wavelength-dependent. In in conventional metasurfaces, different wavelengths will make optical response deviate from the optimized ones, which causes performance degradation. In contrast, our meta-ONN only requires parameters initialized with a Gaussian distribution, without the need for precise values. Therefore, wavelength dependence will not affect the performance of our meta-ONN. This property allows the meta-ONN to operate under the experimental light source with a wide spectral width of 10 nm, which expands its possible applications.
 
\section{Methods}\label{sec11}

\textbf{Metasurface design and simulation} The metasurface is created using massive periodic unit cells. Every unit cell consists of a 220-nm-thick silicon cylindrical meta-atom with a diameter that varies across different cells. The silicon meta-atom is deposited on a 2-$\mu$m-thick buried oxide layer on a 725-$\mu$m-thick silicon substrate. These unit cells are arranged in a square lattice in the two-dimensional plane. The period of the metasurface unit cell is 500 nm.  Based on three-dimensional FDTD simulations, the optical phase and amplitude modulation coefficients of the unit cell for the incident light are obtained at various diameters under normal incidence, see the simulation details in \textcolor{blue}{Supplementary Note 1}. The simulation result indicates that the optical phase and amplitude of the incident light can be effectively manipulated by varying the diameter of the meta-atom, as shown in Figure S1c of \textcolor{blue}{Supplementary Information}. The fabricated metasurface comprises 6400 $\times$ 6400 = 41 million meta-atoms. The diameter of every meta-atom is independently designed to realize the optical phase and amplitude modulation coefficients following a Gaussian sampling distribution. Specifically, the standard deviation of the Gaussian distribution of phase modulation is designed as 0.4 $\pi$. With the designed phase distribution, the diameter of every meta-atom can be obtained according to the mapping relationship between the optical phase and meta-atom diameter as shown in Figure S1c of \textcolor{blue}{Supplementary Information}. The range of the meta-atom diameter is from 100 nm to 400 nm. The amplitude distribution is also determined without the need for further designs according to the relationship between optical amplitude and meta-atom diameter.

\textbf{Metasurface fabrication} The metasurface chip is fabricated based on a multiple-wafer-project (MPW) process offered by a commercial electron-beam-lithography-based silicon foundry. The fabrication process starts with a silicon-on-insulator (SOI) wafer, where the device layer thickness is 220 nm, the buffer oxide layer thickness is 2 $\mu$m, and the silicon substrate thickness is 725 $\mu$m. A layer of electron-sensitive resist is coated on the device layer of the wafer, followed by heating to strengthen the resist. After that, a 100 keV electron gun is used to define the meta-atom patterns of the metasurface. The wafer is then chemically developed, and the patterned resist remains on the substrate. Finally, an anisotropic RIE etching process is performed to transfer the pattern of the resist into the silicon device layer. The etching process will continue until the surface of the buffer oxide layer is exposed.

\textbf{Quantitative phase imaging of the fabricated metasurface} Quantitative Phase Imaging (QPI) is used to measure the reflective phase response of the metasurface. The system has a diffraction-limited spatial resolution of 665 nm and a field of view of 36 × 36 $\mu$m$^2$. With a total magnification of 200 $\times$, the system achieves a high pixel resolution of 60 nm, thus enabling a clear visualization of the phase modulation matrix. The details about the QPI measurement are provided in \textcolor{blue}{Supplementary Note 5}. 

\textbf{Experimental setup of the metasurface-based ONN} Our ONN comprises an SLM for optical image generation, a single-layer metasurface chip, a converging lens for coherent summation, and a camera for optical image capture. The experiment setup starts with a 532-nm laser diode which emits an optical beam with a diameter of 3 mm and a spectrum with a full width at half maximum (FWHM) of 10 nm. The laser is then collimated using an optical lens and expanded to a diameter of 10 mm using an optical 4-$f$ system. The purpose of expanding the beam diameter is to ensure that the laser beam can effectively cover the active area (15.4 mm $\times$ 9.6 mm) of the spatial light modulator (SLM, Meadowlark Optics E19x12-400-800-HDM8). The SLM has a full-pixel resolution of 1,920 $\times$ 1,200, a pitch size of 8 $\mu$m, and an 8-bit modulation depth. The expanded laser beam enters onto the SLM at an off-axis angle of 12°. By introducing two linear polarizers placed before and after the SLM, the SLM can be used to encode the digital image onto the optical domain, as introduced in \textcolor{blue}{Supplementary Note 3}. The polarization angle of both polarizers is 45° with respect to the horizontal direction. The SLM-generated optical image is projected onto the metasurface chip. Here, to ensure that the optical image can effectively overlap with the area of the metasurface chip to realize full-area modulation of the optical image, the effective area of the optical image is intentionally set as 4.1 $\times$ 4.1 mm$^2$. Two approaches are used to realize the consistent effective area in all the tasks: (1) For the task images whose pixel scale is smaller than 512 $\times$ 512, the images are upsampled to a pixel scale of 512 $\times$ 512 using the nearest area interpolation method in the digital domain. Consequently, the effective area of the optical image is 512 $\times$ 8 $\mu$m $\times$ 512 × 8 $\mu$m = 4.1 × 4.1 mm$^2$. For example, the pixel scale of the CIFAR-10 image is upsampled to be 512 $\times$ 512 from 32 $\times$ 32. (2) For the task images whose pixel scale is larger than 512 $\times$ 512, the generated optical image is minified to an effective area of 4.1 $\times$ 4.1 mm$^2$ using a 4-$f$ system in the optical domain. For example, the Chest X-ray image with a pixel scale of 1024 $\times$ 1024 is minified double. In addition, since the effective area of the optical image is smaller than the laser beam size (10 $\times$ 10 mm$^2$), an aperture of 5.5 $\times$ 5.5 mm$^2$ is used to block the stray light of the laser beam. A 50/50 beam splitter is placed between the SLM and the metasurface chip to realize the normal incidence of the optical image to the metasurface. The optical image is reflected by the metasurface and the beam splitter in sequence. Following that, the reflected optical image is focused by an optical converging lens with a focal length of 150 mm. Lastly, a camera (IRVI Contour IR Digital CMOS Camera) with an 8-bit depth and an activated pixel number of 600 $\times$ 800 is placed close to the focal plane of the converging lens to capture the optical image. The figure showing the experimental setup is shown in Figure S3 of \textcolor{blue}{Supplementary Information}.

\textbf{Neural network training} In the experiment, the camera-captured images processed by the meta-ONN are fed into a single-layer backend digital neural network to produce the prediction results of various machine vision tasks. The working flow of neural network training is shown in Figure S9a of \textcolor{blue}{Supplementary Information}. First, the captured images are downsampled by averaging the received power in the neighboring sensor array, see the details in Figure S9b of \textcolor{blue}{Supplementary Information}. After that, the down-sampled images and the corresponding labels are combined as the new dataset after the meta-ONN. The dataset is then randomly split into the training dataset and testing dataset by a certain ratio. The training dataset is used to train the digital neural network. The digital neural network consists of an input layer for receiving the input image and an output layer for generating prediction results, and the two layers are fully connected, as shown in Figure S9c of \textcolor{blue}{Supplementary Information}. Biases are added to the output before the activation function of sigmoid. The mathematical expression of the single-layered digital neural network is $y = sigmoid (b_{0} + w_{0}x)$, where $b_{0}$, $w_{0}$, $x$, and $y$ are the biases, the digital weights, the input image in the form of vector and the prediction result, respectively. Logistic regression is used to optimize the biases and digital weights. Finally, the testing dataset is used to verify the performance of the overall system. The parameters for downsampling the captured images, the split ratio of the dataset, and the hyperparameters for neural network training are task-dependent, which are introduced in \textcolor{blue}{Supplementary Notes 11-17}. The details about the digital model training (SAM, ResNet-50, and ViT) are provided in \textcolor{blue}{Supplementary Note 18}.\\

~\\
~\\
~\\
~\\
~\\
~\\
~\\
~\\
~\\
\\
\\

\backmatter

\begin{appendices}

\definecolor{mycolor_0}{RGB}{214,213,193}
\definecolor{mycolor_1}{RGB}{250,250,246}
\definecolor{mycolor_2}{RGB}{236,235,226}

\begin{flushleft}
\Large \textbf{Appendix A Extended Data}
\end{flushleft}

\begin{table}[htp]

\caption{The performance comparison of the meta-ONN with the state-of-the-art ONN systems}
\renewcommand{\arraystretch}{2}
\begin{tabular}{ccccccc}

\hline

\makecell{\textbf{References} }&\makecell{\textbf{ Methods }} &\makecell{\textbf{Dataset} }&\makecell{\textbf{Processed }\\\textbf{data size$^{2}$}}& \makecell{\textbf{\# of} \\ \textbf{neuron} }& \makecell{\textbf{Task type}} & \makecell{\textbf{Task}\\\textbf{performance}}\\
\hline
\makecell{Huang 2021 [17]\\Shen 2017 [18]\\Ashtiani 2022 [37]\\Feldmann 2021 [38]\\Moralis 2022 [95]} & \makecell{2D integrated \\ photonics} & \makecell{FNC$^{1}$\\Vowel\\EMNIST\\MNIST\\CIFAR-10} & \makecell{1 $\times$ 1 \\ 1 $\times$ 4\\3 $\times$ 4\\28 $\times$ 28\\32 $\times$ 32} & \makecell{8\\16\\9\\16\\7} & \makecell{Classification}& \makecell{N/A \\76.7\%\\89.9\%\\95.3\%\\79.8\%$^{3}$}\\

\hline
\makecell{Lin 2018 [41]\\Wei 2023 [94]\\Zhou 2021 [46]\\Antonik 2019 [77]}& \makecell{3D free-space \\ photonics}&\makecell{MNIST\\CIFAR-10\\KTH\\KTH} &\makecell{28 $\times$ 28\\32 $\times$ 32\\120 $\times$ 160\\120 $\times$ 160} & \makecell{2 $\times$ 10$^5$ \\3.9 $\times$ 10$^4$\\4.9 $\times$ 10$^5$\\1.6 $\times$ 10$^3$} & \makecell{Classification} &\makecell{91.75\%\\73.12\%\\96.3\%\\91.3\%}\\

\hline
\makecell{Teğin 2021 [79]\\Oguz 2022 [76]\\ }& \makecell{Multimode \\ optical fiber} &\makecell{ COVID-19\\COVID-19} &\makecell{299 $\times$ 299\\299 $\times$ 299} & \makecell{N/A\\ N/A} &  \makecell{Classification}&\makecell{83.2\%\\77\%}\\

\hline

\makecell{Luo 2022 [48]\\Qu 2022 [106]\\Zheng 2022 [49]\\ Zheng 2024 [50]}& \makecell{Optical \\ metasurface}&\makecell{ MNIST\\MNIST\\MNIST\\MNIST} &\makecell{28 $\times$ 28\\28 $\times$ 28 \\28 $\times$ 28\\28 $\times$ 28} & \makecell{7.84 $\times$ 10$^4$\\ 1.25 $\times$ 10$^4$\\ 6.27 $\times$ 10$^5$\\ 3.39 $\times$ 10$^5$ } & \makecell{Classification} & \makecell{93.75\%$^{4}$\\98.05\%\\93.1\%\\98.6\%}\\

\hline

\makecell{\textbf{This work}}& \makecell{Optical \\ metasurface}&\makecell{ MNIST\\KTH\\COVID-19\\ Brain ICH\\ChestX-ray8 \\  \makecell{Brain ICH} \\  \makecell{CAMELYON} } &\makecell{28 $\times$ 28\\120 $\times$ 160\\299 $\times$ 299\\512 $\times$ 512\\1,024 $\times$ 1,024\\  \makecell{512 $\times$ 512}\\ \makecell{2 billion pixels$^{5}$}} & \makecell{4.1 $\times$ 10$^7$} & \makecell{~\\~\\Classification ~\\ ~\\ ~\\ Localization \\ Segmentation} & \makecell{99.3\%\\99.1\%\\97.0\%\\ 97.8\%\\85.4\%\\ \makecell{IOU$=$0.61} \\ \makecell{IOU$=$0.60}}\\
\hline

\end{tabular}
\label{Extended_table1}
\footnotetext{$^1$FNC represents the optical communication data for fiber nonlinearity compensation.}
\footnotetext{$^{2}$The processed data size refers to the original data size of the dataset, instead of the data size that can be loaded into the ONN at one time.}
\footnotetext{$^{3}$The accuracy is experimentally obtained based on the two-categorized CIFAR-10 dataset, and the image is preprocessed by an electrical convolutional neural network before being fed into the ONN.}
\footnotetext{$^{4}$The accuracy is experimentally obtained based on the four-categorized MNIST dataset.}
\footnotetext{$^{5}$The whole slide image (WSI) from the CAMELYON-16 dataset has $\sim$ 2 billion pixels, which are cropped into many sub-images with 1,000 $\times$ 1,000 pixels that our system can process in one shot.} 
\footnotetext{\# represents number.}
\end{table}

\end{appendices}
\backmatter

\bmhead{Supplementary information}\label{SI}

Supplementary material is available at submission materials

\bmhead{Acknowledgements}

This work was supported by RGC ECS 24203724, NSFC 62405258, RGC YCRG C4004-24Y, C1002-22Y, ITF ITS/237/22, ITS/226/21FP, RNE-p4-22 of the Shun Hing Institute of Advanced Engineering, NSFC/RGC Joint Research Scheme N$\underline{~}$CUHK444/22.

\bmhead{Author contributions}
C.H. and M.L. conceived the idea. M.L. designed the metasurface chip, conducted the experiments, and analyzed the experimental data. M.L., J.X., and W.G. conducted the optical simulations. M.J. conducted experiments for digital neural networks. N.Z. conducted the QPI experiment. R.Z supervised the QPI experiment. C.H. led the writing of the main manuscript. M.L and B.S, M.J. C.S, and Q.D. contributed to the paper writing. C.H, Q.D., and C.S. supervised the project.

\bmhead{Data availability}
The data that support the plots within this paper and other findings of this study are available from the corresponding author upon reasonable request.

\bmhead{Code availability}
Accession codes will be available before publication.

\bmhead{Competing interest deceleration} Chaoran Huang, Chester Shu, and Mingcheng Luo declare US provisional patent application number 63/643,973.

\newpage
\begin{flushleft}
\large \textbf{References}    
\end{flushleft}
~\\
\noindent [1] LeCun, Y., Bengio, Y., Hinton, G.: Deep learning. nature 521(7553), 436–444 (2015)\\

\noindent [2] Achiam, J., Adler, S., Agarwal, S., Ahmad, L., Akkaya, I., Aleman, F.L., Almeida, D., Altenschmidt, J., Altman, S., Anadkat, S., et al.: Gpt-4 technical report. arXiv preprint arXiv:2303.08774 (2023)\\

\noindent [3] Ahmed, A.H., Sharkia, A., Casper, B., Mirabbasi, S., Shekhar, S.: Siliconphotonics microring links for datacenters—challenges and opportunities. IEEE Journal of Selected Topics in Quantum Electronics 22(6), 194–203 (2016)\\

\noindent [4] Miller, D.A.: Rationale and challenges for optical interconnects to electronic chips. Proceedings of the IEEE 88(6), 728–749 (2000)\\

\noindent [5] Nahmias, M.A., De Lima, T.F., Tait, A.N., Peng, H.-T., Shastri, B.J., Prucnal, P.R.: Photonic multiply-accumulate operations for neural networks. IEEE Journal of Selected Topics in Quantum Electronics 26(1), 1–18 (2019)\\

\noindent [6] Sze, V., Chen, Y.-H., Emer, J., Suleiman, A., Zhang, Z.: Hardware for machine learning: Challenges and opportunities. In: 2017 IEEE Custom Integrated Circuits Conference (CICC), pp. 1–8 (2017). \\

\noindent [7] Shastri, B.J., Tait, A.N., Lima, T., Pernice, W.H., Bhaskaran, H., Wright, C.D., Prucnal, P.R.: Photonics for artificial intelligence and neuromorphic computing. Nature Photonics 15(2), 102–114 (2021)\\

\noindent [8] Markovi´c, D., Mizrahi, A., Querlioz, D., Grollier, J.: Physics for neuromorphic computing. Nature Reviews Physics 2(9), 499–510 (2020)\\

\noindent [9] Huang, C., Sorger, V.J., Miscuglio, M., Al-Qadasi, M., Mukherjee, A., Lampe, L., Nichols, M., Tait, A.N., Lima, T., Marquez, B.A., et al.: Prospects and applications of photonic neural networks. Advances in Physics: X 7(1), 1981155 (2022)\\

\noindent [10] Wetzstein, G., Ozcan, A., Gigan, S., Fan, S., Englund, D., Soljaˇci´c, M., Denz, C., Miller, D.A., Psaltis, D.: Inference in artificial intelligence with deep optics and photonics. Nature 588(7836), 39–47 (2020)\\

\noindent [11] McMahon, P.L.: The physics of optical computing. Nature Reviews Physics 5(12), 717–734 (2023)\\

\noindent [12] Berggren, K., Xia, Q., Likharev, K.K., Strukov, D.B., Jiang, H., Mikolajick, T., Querlioz, D., Salinga, M., Erickson, J.R., Pi, S., et al.: Roadmap on emerging hardware and technology for machine learning. Nanotechnology 32(1), 012002 (2020)\\
\noindent [13] Zhou, H., Dong, J., Cheng, J., Dong, W., Huang, C., Shen, Y., Zhang, Q., Gu,M., Qian, C., Chen, H., et al.: Photonic matrix multiplication lights up photonicaccelerator and beyond. Light: Science \& Applications 11(1), 30 (2022)\\

\noindent [14] Farhat, N.H., Psaltis, D., Prata, A., Paek, E.: Optical implementation of thehopfield model. Applied optics 24(10), 1469–1475 (1985)\\

\noindent [15] Bogaerts, W., P´erez, D., Capmany, J., Miller, D.A., Poon, J., Englund, D.,Morichetti, F., Melloni, A.: Programmable photonic circuits. Nature 586(7828),207–216 (2020)\\

\noindent [16] Tait, A.N., De Lima, T.F., Zhou, E., Wu, A.X., Nahmias, M.A., Shastri, B.J.,Prucnal, P.R.: Neuromorphic photonic networks using silicon photonic weightbanks. Scientific reports 7(1), 7430 (2017)\\

\noindent [17] Huang, C., Fujisawa, S., Lima, T.F., Tait, A.N., Blow, E.C., Tian, Y., Bilodeau,S., Jha, A., Yaman, F., Peng, H.-T., et al.: A silicon photonic–electronic neuralnetwork for fibre nonlinearity compensation. Nature Electronics 4(11), 837–844(2021)\\

\noindent [18] Shen, Y., Harris, N.C., Skirlo, S., Prabhu, M., Baehr-Jones, T., Hochberg, M.,Sun, X., Zhao, S., Larochelle, H., Englund, D., et al.: Deep learning with coherentnanophotonic circuits. Nature photonics 11(7), 441–446 (2017)\\

\noindent [19] Stark, P., Horst, F., Dangel, R., Weiss, J., Offrein, B.J.: Opportunities forintegrated photonic neural networks. Nanophotonics 9(13), 4221–4232 (2020)\\

\noindent [20] Pappas, C., Moschos, T., Moralis-Pegios, M., Giamougiannis, G., Tsakyridis,A., Kirtas, M., Passalis, N., Tefas, A., Pleros, N.: A teraflop photonic matrixmultiplier using time-space-wavelength multiplexed awgr-based architectures.In: 2024 Optical Fiber Communications Conference and Exhibition (OFC), pp.1–3 (2024). IEEE\\

\noindent [21] De Marinis, L., Cococcioni, M., Liboiron-Ladouceur, O., Contestabile, G., Castoldi, P., Andriolli, N.: Photonic integrated reconfigurable linear processors asneural network accelerators. Applied Sciences 11(13), 6232 (2021)\\

\noindent [22] Donati, G., Mirasso, C.R., Mancinelli, M., Pavesi, L., Argyris, A.: Microringresonators with external optical feedback for time delay reservoir computing.Optics Express 30(1), 522–537 (2022)\\

\noindent [23] Skalli, A., Robertson, J., Owen-Newns, D., Hejda, M., Porte, X., Reitzenstein,S., Hurtado, A., Brunner, D.: Photonic neuromorphic computing using verticalcavity semiconductor lasers. Optical Materials Express 12(6), 2395–2414 (2022)\\

\noindent [24] Sozos, K., Bogris, A., Bienstman, P., Sarantoglou, G., Deligiannidis, S., Mesaritakis, C.: High-speed photonic neuromorphic computing using recurrent optical spectrum slicing neural networks. Communications Engineering 1(1), 24 (2022)\\

\setlength{\parindent}{0em}
[25] Amin, R., George, J.K., Wang, H., Maiti, R., Ma, Z., Dalir, H., Khurgin, J.B.,Sorger, V.J.: An ito–graphene heterojunction integrated absorption modulator on si-photonics for neuromorphic nonlinear activation. Apl Photonics 6(12)(2021)\\

[26] Xu, X., Tan, M., Corcoran, B., Wu, J., Boes, A., Nguyen, T.G., Chu, S.T.,Little, B.E., Hicks, D.G., Morandotti, R., et al.: 11 tops photonic convolutionalaccelerator for optical neural networks. Nature 589(7840), 44–51 (2021)\\

[27] Shi, B., Calabretta, N., Stabile, R.: Inp photonic integrated multi-layer neuralnetworks: Architecture and performance analysis. APL Photonics 7(1) (2022)\\

[28] Wu, C., Yu, H., Lee, S., Peng, R., Takeuchi, I., Li, M.: Programmable phasechange metasurfaces on waveguides for multimode photonic convolutional neuralnetwork. Nature communications 12(1), 96 (2021)\\

[29] Dong, B., Aggarwal, S., Zhou, W., Ali, U.E., Farmakidis, N., Lee, J.S., He, Y.,Li, X., Kwong, D.-L., Wright, C., et al.: Higher-dimensional processing usinga photonic tensor core with continuous-time data. Nature Photonics 17(12),1080–1088 (2023)\\

[30] Shainline, J.M., Buckley, S.M., Mirin, R.P., Nam, S.W.: Superconducting optoelectronic circuits for neuromorphic computing. Physical Review Applied 7(3),034013 (2017)\\

[31] Xiao, X., On, M.B., Van Vaerenbergh, T., Liang, D., Beausoleil, R.G., Yoo, S.:Large-scale and energy-efficient tensorized optical neural networks on iii–v-onsilicon moscap platform. Apl Photonics 6(12) (2021)\\

[32] Fu, T., Zang, Y., Huang, Y., Du, Z., Huang, H., Hu, C., Chen, M., Yang, S.,Chen, H.: Photonic machine learning with on-chip diffractive optics. NatureCommunications 14(1), 70 (2023)\\

[33] Cheng, J., Zhao, Y., Zhang, W., Zhou, H., Huang, D., Zhu, Q., Guo, Y., Xu, B.,Dong, J., Zhang, X.: A small microring array that performs large complex-valuedmatrix-vector multiplication. Frontiers of Optoelectronics 15(1), 15 (2022)\\

[34] Bai, B., Yang, Q., Shu, H., Chang, L., Yang, F., Shen, B., Tao, Z., Wang, J.,Xu, S., Xie, W., et al.: Microcomb-based integrated photonic processing unit.Nature Communications 14(1), 66 (2023)
[35] Xu, S., Wang, J., Yi, S., Zou, W.: High-order tensor flow processing usingintegrated photonic circuits. Nature communications 13(1), 7970 (2022)\\

[36] Meng, X., Zhang, G., Shi, N., Li, G., Aza˜na, J., Capmany, J., Yao, J., Shen, Y., Li, W., Zhu, N., et al.: Compact optical convolution processing unit based onmultimode interference. Nature Communications 14(1), 3000 (2023)\\

[37] Ashtiani, F., Geers, A.J., Aflatouni, F.: An on-chip photonic deep neural networkfor image classification. Nature 606(7914), 501–506 (2022)\\

[38] Feldmann, J., Youngblood, N., Karpov, M., Gehring, H., Li, X., Stappers, M.,Le Gallo, M., Fu, X., Lukashchuk, A., Raja, A.S., et al.: Parallel convolutionalprocessing using an integrated photonic tensor core. Nature 589(7840), 52–58(2021)\\

[39] Sun, J., Timurdogan, E., Yaacobi, A., Hosseini, E.S., Watts, M.R.: Large-scalenanophotonic phased array. Nature 493(7431), 195–199 (2013)\\

[40] Zhang, X., Kwon, K., Henriksson, J., Luo, J., Wu, M.C.: A largescale microelectromechanical-systems-based silicon photonics lidar. Nature603(7900), 253–258 (2022)\\

[41] Lin, X., Rivenson, Y., Yardimci, N.T., Veli, M., Luo, Y., Jarrahi, M., Ozcan,A.: All-optical machine learning using diffractive deep neural networks. Science361(6406), 1004–1008 (2018)\\

[42] Dinc, N.U., Saba, A., Madrid-Wolff, J., Gigli, C., Boniface, A., Moser, C., Psaltis,D.: From 3d to 2d and back again. Nanophotonics 12(5), 777–793 (2023)\\

[43] Yildirim, M., Dinc, N.U., Oguz, I., Psaltis, D., Moser, C.: Nonlinear processingwith linear optics. arXiv preprint arXiv:2307.08533 (2023)\\

[44] Wang, T., Sohoni, M.M., Wright, L.G., Stein, M.M., Ma, S.-Y., Onodera,T., Anderson, M.G., McMahon, P.L.: Image sensing with multilayer nonlinearoptical neural networks. Nature Photonics 17(5), 408–415 (2023)\\

[45] Bernstein, L., Sludds, A., Panuski, C., Trajtenberg-Mills, S., Hamerly, R.,Englund, D.: Single-shot optical neural network. Science Advances 9(25), 7904(2023)\\

[46] Zhou, T., Lin, X., Wu, J., Chen, Y., Xie, H., Li, Y., Fan, J., Wu, H., Fang, L., Dai,Q.: Large-scale neuromorphic optoelectronic computing with a reconfigurablediffractive processing unit. Nature Photonics 15(5), 367–373 (2021)\\

[47] Goi, E., Chen, X., Zhang, Q., Cumming, B.P., Schoenhardt, S., Luan, H.,Gu, M.: Nanoprinted high-neuron-density optical linear perceptrons performingnear-infrared inference on a cmos chip. Light: Science \& Applications 10(1), 40(2021)\\

[48] Luo, X., Hu, Y., Ou, X., Li, X., Lai, J., Liu, N., Cheng, X., Pan, A., Duan,H.: Metasurface-enabled on-chip multiplexed diffractive neural networks in thevisible. Light: Science \& Applications 11(1), 158 (2022)\\

[49] Zheng, H., Liu, Q., Zhou, Y., Kravchenko, I.I., Huo, Y., Valentine, J.: Meta-opticaccelerators for object classifiers. Science Advances 8(30), 6410 (2022)\\

[50] Zheng, H., Liu, Q., Kravchenko, I.I., Zhang, X., Huo, Y., Valentine, J.G.: Multichannel meta-imagers for accelerating machine vision. Nature Nanotechnology,1–8 (2024)\\

[51] Vaswani, A., Shazeer, N., Parmar, N., Uszkoreit, J., Jones, L., Gomez, A.N.,Kaiser,  L., Polosukhin, I.: Attention is all you need. Advances in neuralinformation processing systems 30 (2017)\\

[52] Dosovitskiy, A., Beyer, L., Kolesnikov, A., Weissenborn, D., Zhai, X.,Unterthiner, T., Dehghani, M., Minderer, M., Heigold, G., Gelly, S., et al.: Animage is worth 16x16 words: Transformers for image recognition at scale. arXivpreprint arXiv:2010.11929 (2020)\\

[53] Su, V.-C., Chu, C.H., Sun, G., Tsai, D.P.: Advances in optical metasurfaces:fabrication and applications. Optics express 26(10), 13148–13182 (2018)\\

[54] Chen, W.T., Zhu, A.Y., Capasso, F.: Flat optics with dispersion-engineeredmetasurfaces. Nature Reviews Materials 5(8), 604–620 (2020)\\

[55] Neshev, D., Aharonovich, I.: Optical metasurfaces: new generation buildingblocks for multi-functional optics. Light: Science \& Applications 7(1), 58 (2018)\\

[56] Zhao, Y., Liu, X.-X., Al`u, A.: Recent advances on optical metasurfaces. Journalof Optics 16(12), 123001 (2014)\\

[57] Chen, S., Liu, W., Li, Z., Cheng, H., Tian, J.: Metasurface-empowered opticalmultiplexing and multifunction. Advanced Materials 32(3), 1805912 (2020)\\

[58] Liu, F., Huang, X., Chen, Y., Suykens, J.A.: Random features for kernel approximation: A survey on algorithms, theory, and beyond. IEEE Transactions onPattern Analysis and Machine Intelligence 44(10), 7128–7148 (2021)\\

[59] Rahimi, A., Recht, B.: Random features for large-scale kernel machines.Advances in neural information processing systems 20 (2007)\\

[60] Engelberg, J., Zhou, C., Mazurski, N., Bar-David, J., Kristensen, A., Levy, U.:Near-ir wide-field-of-view huygens metalens for outdoor imaging applications.Nanophotonics 9(2), 361–370 (2020)\\

[61] Park, J.-S., Zhang, S., She, A., Chen, W.T., Lin, P., Yousef, K.M., Cheng, J.-X.,Capasso, F.: All-glass, large metalens at visible wavelength using deep-ultravioletprojection lithography. Nano letters 19(12), 8673–8682 (2019)\\

[62] Afridi, A., Canet-Ferrer, J., Philippet, L., Osmond, J., Berto, P., Quidant, R.:Electrically driven varifocal silicon metalens. Acs Photonics 5(11), 4497–4503(2018)\\

[63] Zhou, Z., Li, J., Su, R., Yao, B., Fang, H., Li, K., Zhou, L., Liu, J., Stellinga,D., Reardon, C.P., et al.: Efficient silicon metasurfaces for visible light. AcsPhotonics 4(3), 544–551 (2017)\\

[64] Jacot, A., Gabriel, F., Hongler, C.: Neural tangent kernel: Convergence andgeneralization in neural networks. Advances in neural information processingsystems 31 (2018)\\

[65] Williams, C.: Computing with infinite networks. Advances in neural informationprocessing systems 9 (1996)\\

[66] Yang, G., Hu, E.J.: Feature learning in infinite-width neural networks. arXivpreprint arXiv:2011.14522 (2020)\\

[67] Arora, S., Du, S.S., Hu, W., Li, Z., Salakhutdinov, R.R., Wang, R.: On exactcomputation with an infinitely wide neural net. Advances in neural informationprocessing systems 32 (2019)\\

[68] Arora, S., Du, S., Hu, W., Li, Z., Wang, R.: Fine-grained analysis of optimization and generalization for overparameterized two-layer neural networks. In:International Conference on Machine Learning, pp. 322–332 (2019). PMLR\\

[69] Bietti, A., Mairal, J.: On the inductive bias of neural tangent kernels. Advancesin Neural Information Processing Systems 32 (2019)\\

[70] Tancik, M., Srinivasan, P., Mildenhall, B., Fridovich-Keil, S., Raghavan, N.,Singhal, U., Ramamoorthi, R., Barron, J., Ng, R.: Fourier features let networkslearn high frequency functions in low dimensional domains. Advances in neuralinformation processing systems 33, 7537–7547 (2020)\\

[71] Zhu, W., Yang, R., Geng, G., Fan, Y., Guo, X., Li, P., Fu, Q., Zhang, F., Gu,C., Li, J.: Titanium dioxide metasurface manipulating high-efficiency and broadband photonic spin hall effect in visible regime. Nanophotonics 9(14), 4327–4335(2020)\\

[72] Yang, W., Xiao, S., Song, Q., Liu, Y., Wu, Y., Wang, S., Yu, J., Han, J., Tsai,D.-P.: All-dielectric metasurface for high-performance structural color. Naturecommunications 11(1), 1864 (2020)\\

[73] Devlin, R.C., Khorasaninejad, M., Chen, W.T., Oh, J., Capasso, F.: Broadbandhigh-efficiency dielectric metasurfaces for the visible spectrum. Proceedings ofthe National Academy of Sciences 113(38), 10473–10478 (2016) \\

[74] Saade, A., Caltagirone, F., Carron, I., Daudet, L., Dr´emeau, A., Gigan, S.,Krzakala, F.: Random projections through multiple optical scattering: Approximating kernels at the speed of light. In: 2016 IEEE International Conferenceon Acoustics, Speech and Signal Processing (ICASSP), pp. 6215–6219 (2016).IEEE\\

[75] Xia, F., Kim, K., Eliezer, Y., Han, S., Shaughnessy, L., Gigan, S., Cao, H.: Nonlinear optical encoding enabled by recurrent linear scattering. Nature Photonics18(10), 1067–1075 (2024)\\

[76] Oguz, I., Hsieh, J.-L., Dinc, N.U., Te˘gin, U., Yildirim, M., Gigli, C., Moser, C.,Psaltis, D.: Programming Nonlinear Propagation for Efficient Optical LearningMachines (2022)\\

[77] Antonik, P., Marsal, N., Brunner, D., Rontani, D.: Human action recognitionwith a large-scale brain-inspired photonic computer. Nature Machine Intelligence1(11), 530–537 (2019)\\

[78] Hao, W., Jiaqi, H., Andrea, M., Alfonso, N., Fei, X., Xuanchen, L., Romolo,S., Qiang, L., Rachel, G., Sylvain, G.: Large-scale photonic computing withnonlinear disordered media. Nature Computational Science 4(5), 11–21 (2024)\\

[79] Te˘gin, U., Yıldırım, M., O˘guz, I., Moser, C., Psaltis, D.: Scalable optical learning ˙operator. Nature Computational Science 1(8), 542–549 (2021)\\

[80] Pierangeli, D., Marcucci, G., Conti, C.: Photonic extreme learning machine byfree-space optical propagation. Photonics Research 9(8), 1446–1454 (2021)\\

[81] Sande, G., Brunner, D., Soriano, M.C.: Advances in photonic reservoir computing. Nanophotonics 6(3), 561–576 (2017)\\

[82] Brunner, D., Soriano, M.C., Mirasso, C.R., Fischer, I.: Parallel photonic information processing at gigabyte per second data rates using transient states. Naturecommunications 4(1), 1364 (2013)\\

[83] Appeltant, L., Soriano, M.C., Sande, G., Danckaert, J., Massar, S., Dambre, J.,Schrauwen, B., Mirasso, C.R., Fischer, I.: Information processing using a singledynamical node as complex system. Nature communications 2(1), 468 (2011)\\

[84] Vandoorne, K., Mechet, P., Van Vaerenbergh, T., Fiers, M., Morthier, G.,Verstraeten, D., Schrauwen, B., Dambre, J., Bienstman, P.: Experimentaldemonstration of reservoir computing on a silicon photonics chip. Naturecommunications 5(1), 3541 (2014)\\

[85] Bueno, J., Maktoobi, S., Froehly, L., Fischer, I., Jacquot, M., Larger, L.,Brunner, D.: Reinforcement learning in a large-scale photonic recurrent neuralnetwork. Optica 5(6), 756–760 (2018)\\

[86] Belfer, Y., Geifman, A., Galun, M., Basri, R.: Spectral analysis of the neuraltangent kernel for deep residual networks. Journal of Machine Learning Research25(184), 1–49 (2024)\\

[87] Muminov, B., Vuong, L.T.: Fourier optical preprocessing in lieu of deep learning.Optica 7(9), 1079–1088 (2020)\\

[88] Mironovova, M., B´ıla, J.: Fast fourier transform for feature extraction and neuralnetwork for classification of electrocardiogram signals. In: 2015 Fourth International Conference on Future Generation Communication Technology (FGCT),pp. 1–6 (2015). IEEE\\

[89] He, K., Zhang, X., Ren, S., Sun, J.: Deep residual learning for image recognition. In: Proceedings of the IEEE Conference on Computer Vision and PatternRecognition, pp. 770–778 (2016)\\

[90] Kirillov, A., Mintun, E., Ravi, N., Mao, H., Rolland, C., Gustafson, L., Xiao, T.,Whitehead, S., Berg, A.C., Lo, W.-Y., et al.: Segment anything. In: Proceedingsof the IEEE/CVF International Conference on Computer Vision, pp. 4015–4026(2023)\\

[91] Rahman, T., Khandakar, A., Qiblawey, Y., Tahir, A., Kiranyaz, S., Kashem,S.B.A., Islam, M.T., Al Maadeed, S., Zughaier, S.M., Khan, M.S., et al.: Exploring the effect of image enhancement techniques on covid-19 detection using chestx-ray images. Computers in biology and medicine 132, 104319 (2021)\\

[92] Wang, X., Peng, Y., Lu, L., Lu, Z., Bagheri, M., Summers, R.M.: Chestx-ray8:Hospital-scale chest x-ray database and benchmarks on weakly-supervised classification and localization of common thorax diseases. In: Proceedings of theIEEE Conference on Computer Vision and Pattern Recognition, pp. 2097–2106(2017)\\

[93] Flanders, A.E., Prevedello, L.M., Shih, G., Halabi, S.S., Kalpathy-Cramer, J.,Ball, R., Mongan, J.T., Stein, A., Kitamura, F.C., Lungren, M.P., et al.: Construction of a machine learning dataset through collaboration: the rsna 2019brain ct hemorrhage challenge. Radiology: Artificial Intelligence 2(3), 190211(2020)\\

[94] Wei, K., Li, X., Froech, J., Chakravarthula, P., Whitehead, J., Tseng, E., Majumdar, A., Heide, F.: Spatially varying nanophotonic neural networks. ScienceAdvances 10(45), 0391 (2024)\\

[95] Moralis-Pegios, M., Mourgias-Alexandris, G., Tsakyridis, A., Giamougiannis,G., Totovic, A., Dabos, G., Passalis, N., Kirtas, M., Rutirawut, T., Gardes,F., et al.: Neuromorphic silicon photonics and hardware-aware deep learningfor high-speed inference. Journal of Lightwave Technology 40(10), 3243–3254(2022)\\

[96] Schuldt, C., Laptev, I., Caputo, B.: Recognizing human actions: a local svmapproach. In: Proceedings of the 17th International Conference on PatternRecognition, 2004. ICPR 2004., vol. 3, pp. 32–36 (2004). IEEE\\

[97] Gilbert, A., Illingworth, J., Bowden, R.: Action recognition using mined hierarchical compound features. IEEE Transactions on Pattern Analysis and MachineIntelligence 33(5), 883–897 (2010)
[98] Grushin, A., Monner, D.D., Reggia, J.A., Mishra, A.: Robust human actionrecognition via long short-term memory. In: The 2013 International JointConference on Neural Networks (IJCNN), pp. 1–8 (2013). IEEE\\

[99] Campanella, G., Hanna, M.G., Geneslaw, L., Miraflor, A., Werneck Krauss Silva,V., Busam, K.J., Brogi, E., Reuter, V.E., Klimstra, D.S., Fuchs, T.J.: Clinicalgrade computational pathology using weakly supervised deep learning on wholeslide images. Nature medicine 25(8), 1301–1309 (2019)\\

[100] Bejnordi, B.E., Veta, M., Van Diest, P.J., Van Ginneken, B., Karssemeijer, N.,Litjens, G., Van Der Laak, J.A., Hermsen, M., Manson, Q.F., Balkenhol, M.,et al.: Diagnostic assessment of deep learning algorithms for detection of lymphnode metastases in women with breast cancer. Jama 318(22), 2199–2210 (2017)\\

[101] Otsu, N., et al.: A threshold selection method from gray-level histograms.Automatica 11(285-296), 23–27 (1975)\\

[102] Chen, Z., Sludds, A., Davis III, R., Christen, I., Bernstein, L., Ateshian, L.,Heuser, T., Heermeier, N., Lott, J.A., Reitzenstein, S., et al.: Deep learning withcoherent vcsel neural networks. Nature Photonics 17(8), 723–730 (2023)\\

[103] Zhou, F., Chai, Y.: Near-sensor and in-sensor computing. Nature Electronics3(11), 664–671 (2020)\\

[104] Chen, Y., Nazhamaiti, M., Xu, H., Meng, Y., Zhou, T., Li, G., Fan, J., Wei, Q.,Wu, J., Qiao, F., et al.: All-analog photoelectronic chip for high-speed visiontasks. Nature 623(7985), 48–57 (2023)\\

[105] Paniagua-Dominguez, R., Yu, Y.F., Khaidarov, E., Choi, S., Leong, V., Bakker,R.M., Liang, X., Fu, Y.H., Valuckas, V., Krivitsky, L.A., et al.: A metalens witha near-unity numerical aperture. Nano letters 18(3), 2124–2132 (2018)\\

[106] Qu, G., Cai, G., Sha, X., Chen, Q., Cheng, J., Zhang, Y., Han, J., Song, Q.,Xiao, S.: All-dielectric metasurface empowered optical-electronic hybrid neuralnetworks. Laser \& Photonics Reviews 16(10), 2100732 (2022)\\

\end{document}